\documentclass[a4paper]{raa_arxiv}            

\usepackage{graphicx,times}             
\usepackage{natbib}
\usepackage{amssymb,amsmath}
\bibpunct{(}{)}{;}{a}{}{,}

\usepackage[breaklinks,colorlinks,urlcolor=blue,citecolor=blue,linkcolor=blue]{hyperref}
\hypersetup{colorlinks = true, linkcolor = green, anchorcolor = red, citecolor = blue, filecolor = red, pagecolor = red, urlcolor = red}
\newcommand*{\rom}[1]{\uppercase\expandafter{\romannumeral #1\relax}}

\begin{document}

   \title{High--Resolution Observations of Prominence Plume Formation with the New Vacuum Solar Telescope}

   \volnopage{Vol.0 (20xx) No.0, 000--000}      
   \setcounter{page}{1}          

   \author{Jian-Chao Xue
      \inst{1,2}
   \and Jean-Claude Vial
      \inst{3}
   \and Yang Su
      \inst{1,2}
    \and Hui Li
      \inst{1,2}
    \and Zhi Xu
      \inst{4}
    \and Ying-Na Su
      \inst{1,2}
    \and Tuan-Hui Zhou
      \inst{1}
    \and Zhen-Tong Li
      \inst{1,2}      
   }

   \institute{Key Laboratory of Dark Matter and Space Astronomy, Purple Mountain Observatory, Chinese Academy of Sciences, Nanjing 210023, China; {\it yang.su@pmo.ac.cn, nj.lihui@pmo.ac.cn}\\
        \and
             School of Astronomy and Space Science, University of Science and Technology of China, Hefei 230026, China\\
        \and
             Institut d’Astrophysique Spatiale, CNRS/Universit\'e Paris-Sud, Universit\'e Paris-Saclay, B\^atiment 121, 91405 Orsay Cedex, France\\
        \and
             Yunnan Observatories, Chinese Academy of Sciences, Kunming 650216, China\\
\vs\no
   {\small Received~~20xx month day; accepted~~20xx~~month day}}
\abstract{Prominence plumes are evacuated upflows that emerge from bubbles below prominences, whose formation mechanism is still unclear. Here we present a detailed study of plumes in a quiescent prominence using the high--resolution H$\mathrm{\alpha}$ filtergrams at the line center as well as line wing at $\pm 0.4\,\mathrm{\AA}$ from the New Vacuum Solar Telescope. Enhancements of brightening, blue shifts, and turbulence at the fronts of plumes are found during their formation. Some large plumes split at their heads and finger--shaped structures are formed between them. Blue--shifted flows along the bubble--prominence interface are found before and during the plume formation. Our observations are consistent with the hypothesis that prominence plumes are related to coupled Kelvin--Helmholtz and Rayleigh--Taylor (KH/RT) instabilities. Plume splittings and fingers are evidence of RT instability, and the flows may increase the growth rate of KH/RT instabilities. However, the significant turbulence at plume fronts may suggest that the RT instability is triggered by the plumes penetrating into the prominence. In this scenario, extra mechanisms are necessary to drive the plumes.
}

   \authorrunning{J.-C. Xue et al.}            
   \titlerunning{Prominence Plume Formation}  

   \maketitle

%
%
\section{Introduction}           
\label{sect:intro}

Solar prominences are cool and dense structures suspended in the hot and tenuous corona \citep{2015ASSL..415.....V}. When prominences are observed at the solar limb, they are bright in chromospheric (H$\mathrm{\alpha}$, Ca~\rom{2} bands, imaging the prominence core) and transition region lines (He~\rom{2} $304\,\mathrm{\AA}$, Fe~\rom{8} $131\,\mathrm{\AA}$, Fe~\rom{9}  $171\,\mathrm{\AA}$, imaging the prominence--corona transition region (PCTR)), but dark in some extreme ultraviolet (EUV) filtergrams due to continuum photoionization \citep{2010SSRv..151..243L}. With the development of ground-- and space--based telescopes, high--resolution observations are revealing more dynamic motions in prominences and filaments. 

Bubble--like cavities are sometimes observed between the solar limb and quiescent prominences at high latitudes \citep{2010ApJ...716.1288B,2011Natur.472..197B,2012ApJ...761....9D,2015ApJ...814L..17S,2018ApJ...863..192L}. They are darker when observed at low chromospheric temperature  but brighter in some EUV filtergrams than prominences. A prominence bubble generally has a semi-circular shape and the bubble--prominence interface is brighter than the ambient prominence. The interface is sometimes arched upwards and a rising plume is formed \citep{2008ApJ...676L..89B}. Plumes have the maximum speed of $20-30\,\mathrm{km\,s^{-1}}$ with turbulent flows \citep{2010ApJ...716.1288B,2019FrP.....7..218A}. After the disappearance of plumes, prominences almost return to their initial states, until another occurrence of plumes. Plumes occur intermittently without clear spatial or temporal regularity \citep{2010ApJ...716.1288B}.

Both the nature of prominence bubbles and the formation mechanism of plumes are under debate. Some authors proposed that prominence bubbles are emergent flux ropes \citep{2011Natur.472..197B} and suffer buoyancy forces \citep{2008ApJ...676L..89B,2017ApJ...850...60B}. By observations and simulations, plumes are explained as the results of Rayleigh--Taylor (RT) instability due to density inversion \citep{2010ApJ...716.1288B,2011Natur.472..197B,2011ApJ...736L...1H,2015ApJ...806L..13K,2016ApJ...825L..29X}. \citet{2017ApJ...850...60B} found shear flows on bubble boundary and proved in theory that shear flows could increase the growth rate of the coupled Kelvin–Helmholtz and Rayleigh–Taylor (KH/RT) instabilities, hence be conducive to the plume occurrence. RT instability has been successful in explaining the plume observations \citep{2018RvMPP...2....1H}, except that magnetohydrodynamic (MHD) simulations predicted the drop of prominence mass, although the observed prominence bodies are relatively stable. Others proposed that bubbles are bright in EUV images because they are empty, which means less absorption off the background emission \citep{2012ApJ...761....9D,2014A&A...567A.123G}. They included parasitic magnetic bipoles below the force--free models of prominences which results in the arcade structures that are similar to the prominence bubbles. The authors suspected that plumes are caused by magnetic reconnection at the separators between bubbles and prominences. Such kind of magnetic reconnection was used to explain the collapse of bubble boundary and the following downflow and upflow of prominence knots \citep{2015ApJ...814L..17S}. However, during the plume formation, no observation shows a clear connection change at bubble boundary or bidirectional outflows. Besides, \citet{2019FrP.....7..218A} observed rotation--like motions within plumes, and they suspected that it indicates the flux rope configuration of plumes or kink instability in the prominence.

We observed a quiescent prominence on 2018 November 10 using the New Vacuum Solar Telescope \citep[NVST, ][]{2014RAA....14..705L}. High spatial resolution and high quality H$\mathrm{\alpha}$ images recorded the formation processes and rich dynamic motions of plumes. We derive Doppler velocity and nonthermal velocity of prominence plasma from the H$\mathrm{\alpha}$ spectral images, which reveal common features of plumes that are important in understanding their formation. The data reduction and a 3-points method for spectral parameter measurements are introduced in Section~\ref{sect:meth}. The observational results are shown in Section~\ref{sect:res}. The formation mechanisms of prominence plumes are discussed in Section~\ref{sect:dis}, which is followed by the conclusion.

\section{Methods}
\label{sect:meth}

\begin{figure}
   \centering
   \includegraphics[width=5.7in, angle=0]{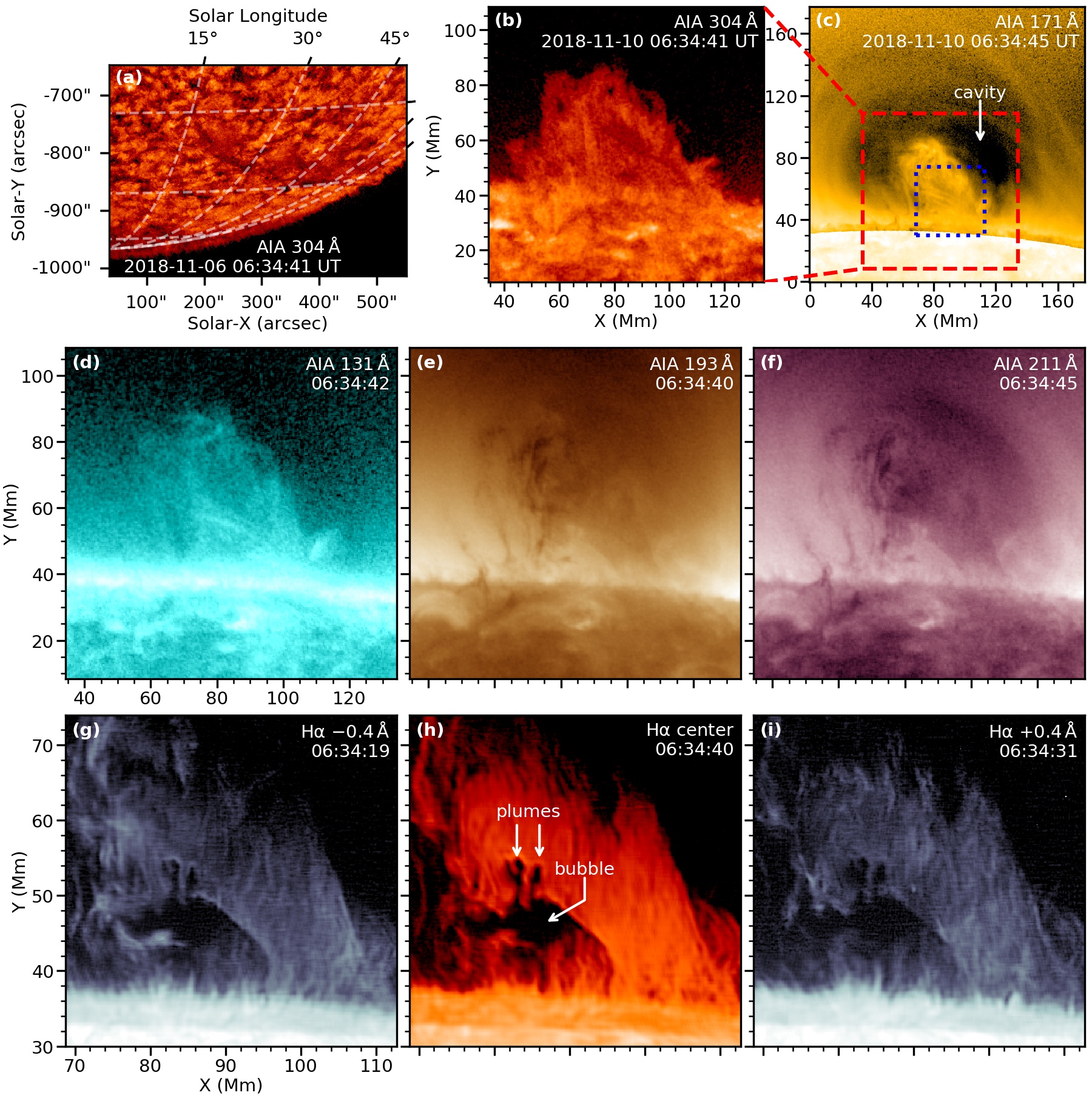}
   \caption{\label{fig:overview}
   Overview of the observations. (a) AIA $304\,\mathrm{\AA}$ map four days before (2018 November 6). (b)--(f) AIA EUV maps of the studied prominence. (g)--(i) NVST H$\mathrm{\alpha}$ maps. The observation channels and times are noted in each panel. The AIA $171\,\mathrm{\AA}$ map in (c) is processed with \emph{aia\_rfilter.pro} to enhance the off--disk emission. The \emph{dashed red square} marks the field of view (FOV) of (b) and (d)--(f), and the \emph{dotted blue square} marks the FOV of (g)--(i). Maps in (b)--(i) are shown in logarithmic scale.}
\end{figure}


The prominence was observed in H$\mathrm{\alpha}$ line center $6562.8\,\mathrm{\AA}$ and off--band $\pm 0.4\,\mathrm{\AA}$ using the NVST. The corresponding EUV images from the Solar Dynamics Observatory \citep{2012SoPh..275....3P}/Atmospheric Imaging Assembly \citep[AIA, ][]{2012SoPh..275...17L} are available (Fig.~\ref{fig:overview}). The H$\mathrm{\alpha}$ spectral images are taken with spatial resolution of $0.136''$ ($98\,\mathrm{km}$) per pixel, passband of $0.25\,\mathrm{\AA}$, exposure time of $20\,\mathrm{ms}$, and a cadence of around 30 seconds. The H$\mathrm{\alpha}$ $+0.4\,\mathrm{\AA}$ maps are multiplied by 1.1 to remove the difference between average $\pm 0.4\,\mathrm{\AA}$ intensities in solar quiet region. The NVST maps are sampled to coalign with each other and with the AIA $193\,\mathrm{\AA}$ maps manually. Other NVST maps are coaligned using our new code Fourier Local Correlation Alignment (\href{https://github.com/xuejcak/flca}{FLCA}). FLCA was developed mainly based on the Fourier Local Correlation Tracking \citep[FLCT, ][]{2008ASPC..383..373F}. It can calculate the correlation between two images at each pixel, and offers a statistical result for image shift. The full-disk AIA EUV maps have a pixel scale of $0.6''$ and a cadence of 12~seconds.  They are processed to level 1.5 before a further use. To improve the signal to noise ratio, five AIA images within 1 minute are averaged. 

To derive Doppler velocity and nonthermal velocity from the NVST H$\mathrm{\alpha}$ observations, we propose a measurement method at three wavelength points. The following sub-sections will introduce this method, its limitations, and comparisons with Gaussian fitting.

\subsection{Spectral parameter measurements using 3-points method}

\citet{1982A&A...115..367S} described the measurement methods of Doppler velocities and line broadening at two and four wavelengths in the same emission line. Taking into account the non-negligible errors, we propose a tentative determination of the three main line parameters at three wavelengths, named as 3-points method.  We assume that the off--disk H$\mathrm{\alpha}$ lines are optically thin and the continua are negligible \citep{1993A&AS...99..513G}. This assumption is possibly satisfied over $6\,\mathrm{Mm}$ in altitude, where H$\mathrm{\alpha}$ center intensities are generally higher than the average intensities of $\pm 0.4\,\mathrm{\AA}$, which means that the integrated opacity is not large \citep{1993A&AS...99..513G, 2014A&A...562A.103H,2015ApJ...800L..13H}. After subtracting the observed stray light, we assume that off--disk H$\mathrm{\alpha}$ lines have Gaussian profiles
\begin{equation}\label{eq:gauss}
I = I_0 \exp\left[-\left(\frac{\lambda-\lambda_\mathrm{D}}{w}\right)^2\right] \, ,
\end{equation}
where $I$ is the intensity at wavelength $\lambda$, $I_0$ is the central intensity, $\lambda_\mathrm{D}$ is the central wavelength modified by a Doppler shift, and $w$ is the Gaussian width. With three off--band H$\mathrm{\alpha}$ observations, plugging wavelengths $\lambda_1-\lambda_3$ and corresponding intensities $I_1-I_3$ into Eq.~(\ref{eq:gauss}) yields
\begin{equation} \label{eq:3pmethod}
\left\{
\begin{array}{l}
\lambda_\mathrm{D}=\frac{1}{2}\frac{\lambda_1^2\ln\frac{I_3}{I_2}+\lambda_2^2\ln\frac{I_1}{I_3}+\lambda_3^2\ln\frac{I_2}{I_1}}{\lambda_1\ln\frac{I_3}{I_2}+\lambda_2\ln\frac{I_1}{I_3}+\lambda_3\ln\frac{I_2}{I_1}}\,,\\
w^2=\frac{\lambda_1^2+2(\lambda_2-\lambda_1)\lambda_\mathrm{D}-\lambda_2^2}{\ln\frac{I_2}{I_1}}\,,\\
I_0=I_1\exp\left[\left(\frac{\lambda_1-\lambda_\mathrm{D}}{w}\right)^2\right]\,.
\end{array} \right.
\end{equation}

The Doppler speed $v_\mathrm{D}$ is derived from
\begin{equation}
v_\mathrm{D}=c\frac{\Delta\lambda_\mathrm{D}}{\lambda_0}=c\frac{\lambda_\mathrm{D}-\lambda_0}{\lambda_0}\,,
\end{equation}
where $c$ is the light speed, $\Delta\lambda_\mathrm{D}$ is wavelength shift due to the Doppler effect, and $\lambda_0=6562.8\,\mathrm{\AA}$ is the H$\mathrm{\alpha}$ rest wavelength. 

If natural broadening and collisional broadening are ignored, the Gaussian width $w$ is mainly contributed by thermal, nonthermal and instrumental broadening:
\begin{equation}
w^2=\left(\frac{\lambda_0}{c}\right)^2(v_\mathrm{t}^2+v_\mathrm{nt}^2)+w_\mathrm{instr}^2\,,
\end{equation}
where $w_\mathrm{instr}$ is the instrumental broadening. The thermal velocity $v_\mathrm{t}$ is defined as:
\begin{equation}
v_\mathrm{t}=\sqrt{\frac{2k_\mathrm{B}T}{m_\mathrm{H}}}\,,
\end{equation}
$k_\mathrm{B}$, $T$, and $m_\mathrm{H}$ are the Boltzmann constant, temperature, and hydrogen mass, respectively. Because $w_\mathrm{instr}$ is unknown, we define $v_\mathrm{nt}$ as:
\begin{equation}
v_\mathrm{nt}=\sqrt{\left(\frac{cw}{\lambda_0}\right)^2-\frac{2k_\mathrm{B}T}{m_\mathrm{H}}}\,,
\end{equation}
which actually includes the instrumental broadening. In Section~\ref{subs:tur}, we will find that $(c/\lambda_0)w_\mathrm{instr}<10\,\mathrm{km\,s^{-1}}$. $v_\mathrm{nt}$ maps are calculated with the assumption $T=9,500\,\mathrm{K}$ for all the off--disk structures \citep{chae_park_song_2013}. $v_\mathrm{nt}$ is generally used to evaluate the plasma turbulence along line of sight (LOS).

\subsection{Limitations of the 3-points method}

\begin{figure}
   \centering
   \includegraphics[width=5.7in, angle=0]{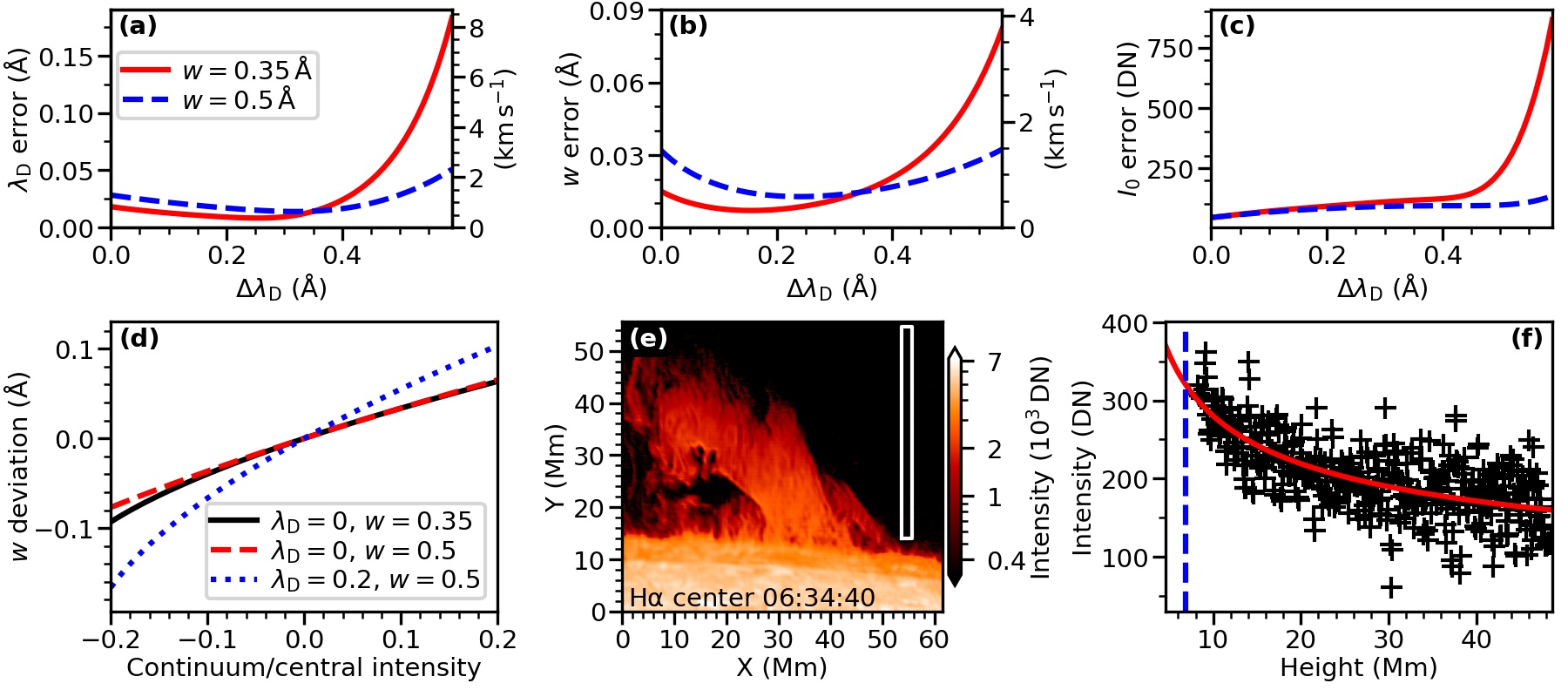}
   \caption{\label{fig:gau}
   Errors of spectral parameter measurements and stray light estimation. (a)--(c) Errors of central wavelength $\lambda_\mathrm{D}$, Gaussian width $w$, and central intensity $I_\mathrm{0}$ versus Doppler shift. Peak intensities of Gaussian profiles are set to be $2,000\,\mathrm{DN}$, and profile widths are noted in (a). (d) Deviation of calculated $w$ from assumed $w$ due to overestimation (negative X-axis values) or underestimation (positive values) of stray light. (e) NVST H$\mathrm{\alpha}$ center map. The \emph{white box} is the region for stray light estimation. (f) \emph{Black pluses}: intensities in the white box in (e) distributed along height above the solar limb; \emph{red curve}: fitting the pluses using a power law; \emph{blue line}: the height above which the spectral parameters are derived.}
\end{figure}

The square root of photon number or digital number (DN) is generally treated as intensity error, which assumes that received photons follow a Poisson distribution. Then we can derive the errors of $\lambda_\mathrm{D}$, $w$, and $I_\mathrm{0}$ using the error propagation formula. For example, error of $\lambda_\mathrm{D}$ is calculated using
\begin{equation} \label{eq:errorlam}
    E\lambda_\mathrm{D} = \sqrt{\left(\frac{\partial \lambda_\mathrm{D}}{\partial I_1}\right )^2 I_1+\left(\frac{\partial \lambda_\mathrm{D}}{\partial I_2}\right )^2 I_2 + \left(\frac{\partial \lambda_\mathrm{D}}{\partial I_3}\right )^2 I_3} \, ,
\end{equation}
where $\frac{\partial \lambda_\mathrm{D}}{\partial I_j}~(j=1,2,3)$ are derived from Eq.~(\ref{eq:3pmethod}).
Since we derive the Gaussian parameters at only three wavelengths, the measurement errors are sensitive to Doppler shifts. To evaluate the variation of $\lambda_\mathrm{D}$, $w$, and $I_\mathrm{0}$ errors with respect to the Doppler shifts $\Delta\lambda_\mathrm{D}$, we set a series of Gaussian profiles with $I_\mathrm{0}=2000\,\mathrm{DN}$ and $w=0.35, 0.5\,\mathrm{\AA}$ according to our observations. With varying $\Delta\lambda_\mathrm{D}$, Gaussian profiles and $I_1-I_3$ are determined, then we can calculate the errors of $\lambda_\mathrm{D}$ (using Eq.~(\ref{eq:errorlam})), $w$, and $I_\mathrm{0}$. The results are plotted in Figs.~\ref{fig:gau}(a)--(c). They show that all the parameter errors vary gently when $\Delta\lambda_\mathrm{D}<0.4\,\mathrm{\AA}$ and vary faster beyond it, which is due to the fact that the chosen wavelengths are within $0.4\,\mathrm{\AA}$. Besides, the wider profile is less sensitive to Doppler shifts than the thinner one. Note that $0.4\,\mathrm{\AA}$ corresponds to a shift of $\sim 18\,\mathrm{km\,s^{-1}}$, and $w=0.35, 0.5\,\mathrm{\AA}$ correspond to $v_\mathrm{nt}=10, 19\,\mathrm{km\,s^{-1}}$, respectively, when $T=9500\,\mathrm{K}$ for H$\mathrm{\alpha}$ line.
This method does not allow any degree of freedom in measurement, so we cannot evaluate how much the observed H$\mathrm{\alpha}$ lines deviate from Gaussian profiles.

The reliability of measured $w$ and $I_\mathrm{0}$ also depends on the estimation of stray light, which is mainly due to the scattering of solar disk emission by Earth's atmosphere and the instrument. Figure~\ref{fig:gau}(d) shows the deviation of measured $w$ from the assumed values (noted in the legend in the units of $\mathrm{\AA}$) versus continuum. Negative continuum value means that stray light is overestimated, in which case $w$ is underestimated, and vice versa. This effect gets more significant when there is a Doppler shift (dotted blue curve). In this work, the stray light is evaluated at a clear region beyond the prominence (the white box in Fig.~\ref{fig:gau}(e)). The stray light is a function of height (plus signs in Fig.~\ref{fig:gau}(f)), and its distribution is fitted by a power law with a negative power index (solid red curve). The calculation of spectral parameters is performed on bright H$\mathrm{\alpha}$ structures above $6.9\,\mathrm{Mm}$ from the solar limb (dashed blue line). In this work, we estimate the intensity error from two components: one is the square root of digital number, the other is the standard deviation of stray light after subtracting the fitting curve.

\subsection{Comparison with Gaussian fitting}

\begin{figure}
   \centering
   \includegraphics[width=5.7in, angle=0]{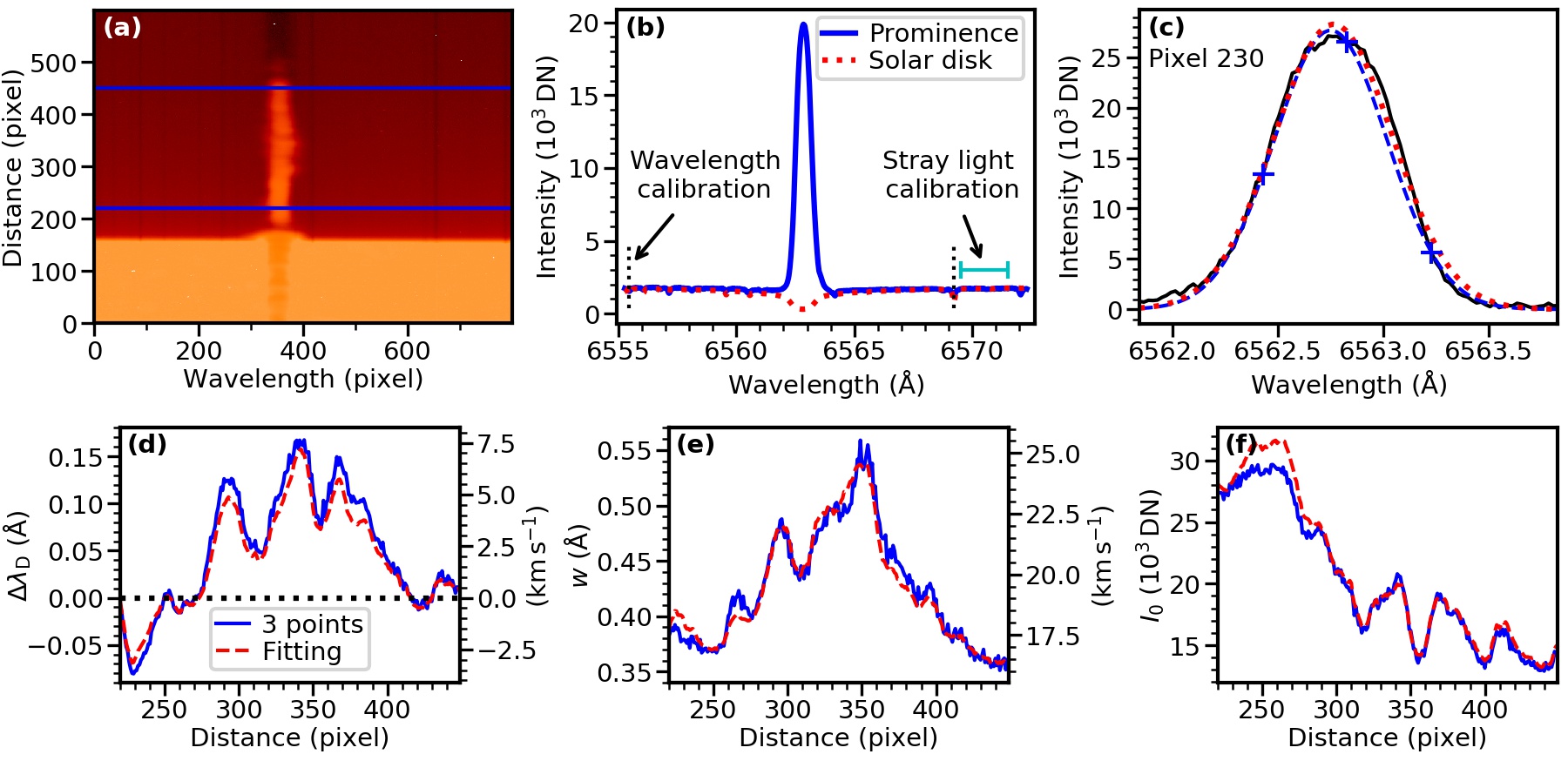}
   \caption{\label{fig:fit}
   Comparison of the 3-points method with Gaussian fitting. (a) H$\mathrm{\alpha}$ spectral observation of another prominence. (b) Comparison of spectra between the prominence and solar disk center. The solid blue curve in (b) is an average spectrum between the \emph{blue lines} in (a); the \emph{red dotted curve} is FTS spectrum multiplied by a coefficient, which is determined by the wavelength range marked with \emph{cyan lines} for stray light calibration; the two faint reference lines with \emph{dotted black lines} are Si~I~$6555.5\,\mathrm{\AA}$ and Fe~I~$6569.2\,\mathrm{\AA}$, respectively. (c) The \emph{solid black curve} is H$\mathrm{\alpha}$ line at distance pixel 230 with stray light subtracted; 0 and $\pm 0.4\,\mathrm{\AA}$ are marked with \emph{blue plus symbols}, and the \emph{dashed blue curve} is the Gaussian profile using the 3-points method; the \emph{dotted red curve} is the result of Gaussian fitting. (d)--(f) Distributions of $\Delta \lambda_\mathrm{D}$, $w$, and $I_\mathrm{0}$ along distance, the \emph{solid blue curves} are results of the 3-points method, and the \emph{dashed red curves} are results of Gaussian fitting.}
\end{figure}

Gaussian fitting is widely used to derive spectral parameters for emission lines. To test the reliability of the 3-points method, we compared it with Gaussian fitting using a set of H$\mathrm{\alpha}$ spectral data of a prominence observed with the Multi--wavelength Spectrometer on the NVST \citep{2013RAA....13.1240W}. Figure~\ref{fig:fit}(a) shows the spectral image with X-axis wavelength and Y-axis distance in units of pixel. The region between two blue lines is taken to make the average H$\mathrm{\alpha}$ profile as shown in Fig.~\ref{fig:fit}(b). Before performing Gaussian parameter measurements on the data, wavelength and stray light calibrations are necessary. The faint absorption lines on both sides of the H$\mathrm{\alpha}$ profile result from the scattering of solar disk light, by which we can determine the wavelength. We adopt the spectral data observed by the Fourier Transform Spectrometer (FTS) at the McMath/Pierce Solar Telescope \citep{1978fsoo.conf...33B} as standard solar disk spectrum (dotted red curve in Fig.~\ref{fig:fit}(b)), and compare them with the average prominence spectrum (solid black curve). The two reference lines for wavelength calibration are marked with vertical dotted lines, which are Si~I~$6555.5\,\mathrm{\AA}$ and Fe~I~$6569.2\,\mathrm{\AA}$, respectively. Intensity of stray light is a function of height and wavelength, and is related to solar disk radiation:
\begin{equation} \label{eq:sl}
    I_\mathrm{SL}(d,\lambda) \approx C(d)I_\mathrm{SD}(d,\lambda) \, ,
\end{equation}
where $I_\mathrm{SL}$ is stray light intensity of observed prominence spectrum, $I_\mathrm{SD}$ represents the solar disk radiation, $d$ is height. The coefficient $C(d)$ is determined using the wavelength range marked with cyan lines, where we assume that the observed prominence radiation is from scattering of solar disk light totally. In Fig.~\ref{fig:fit}(b), the dotted red curve is FTS data multiplied by the coefficient, hence it represents the stray light intensity of the prominence spectrum.

After wavelength calibration and stray light being subtracted, the solid black curve in Fig.~\ref{fig:fit}(c) is the prominence H$\mathrm{\alpha}$ line at distance 230 pixel, as an example. We didn't plot error bars because most errors are too small to show. The 3 plus symbols mark the points at 0 and $\pm 0.4\,\mathrm{\AA}$, and the dashed blue curve is the result of the 3-points method. The dotted red curve is the result of Gaussian fitting using all the observed points. The two calculated profiles are similar. The derived $\Delta \lambda_\mathrm{D}$, $w$, and $I_0$ along distance are shown in Figs.~\ref{fig:fit}(d)--(f), respectively, where solid blue curves are the results of the 3-points method, and dashed red curves are that of Gaussian fitting. The maximum difference of $\Delta \lambda_\mathrm{D}$ of the two methods is $0.031\,\mathrm{\AA}$ (Doppler velocity $\sim 1.4\,\mathrm{km\,s^{-1}}$); that of $w$ is $0.026\,\mathrm{\AA}$ (relative difference 5.9\%); that of $I_0$ is $826\,\mathrm{DN}$ (relative difference 4.1\%).

\section{Observational results} \label{sect:res}
\subsection{Overview}
\begin{figure}
   \centering
   \includegraphics[width=5.7in, angle=0]{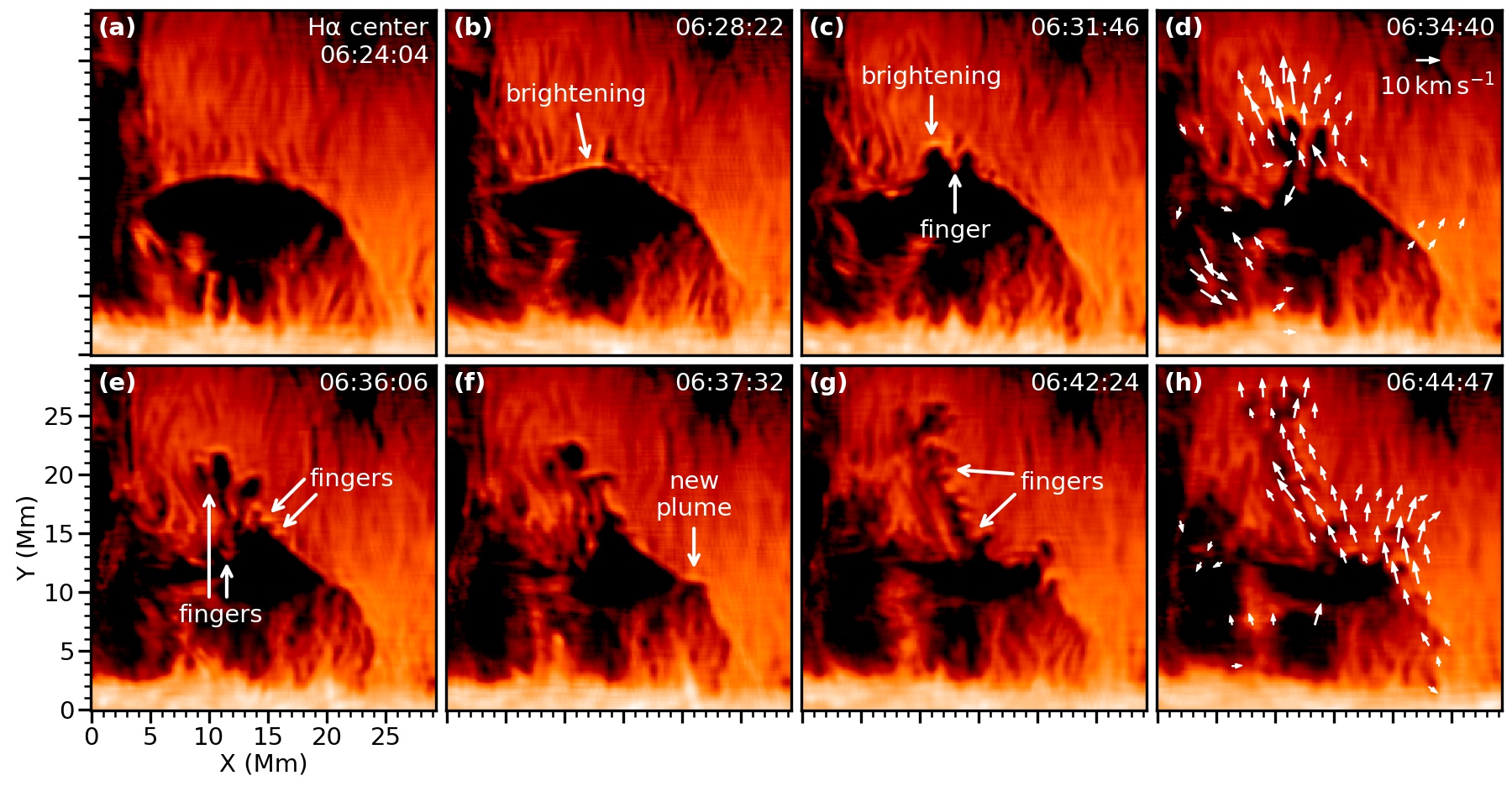}
   \caption{\label{fig:evol}
   Formation and evolution processes of plumes in H$\mathrm{\alpha}$ center in logarithmic scale. Maps in (d) and (h) are overlaid with POS velocities; the velocities $<4\,\mathrm{km\,s^{-1}}$ are not shown.}
\end{figure}
A knowledge of the three--dimensional structure is important to understand the two--dimensional projection of the prominence \citep{2018ApJ...867..115G}. Four days before the observation, the target prominence was seen as a dark filament that had the orientation of northeast--southwest (NE--SW) and east--west (E--W) on the two sides of the longitude $45^\circ$ (Fig.~\ref{fig:overview}(a)). During the four days, the Sun has rotated about $53^\circ$, thus the prominence observed on 2018 November 10 is mainly oriented NE--SW, and part of E--W oriented prominence is blocked by the NE--SW part (Fig.~\ref{fig:overview}(b)). The prominence is embedded in a coronal cavity and large--scale bright loops in AIA $171\,\mathrm{\AA}$ filtergram ($\sim 0.8\,\mathrm{MK}$, Fig.~\ref{fig:overview}(c)). Different from the usual ``tennis racquet'' shaped coronal cavities \citep{2011Natur.472..197B,2018LRSP...15....7G}, the cavity in our observations is relatively wide and low. The prominence is bright in AIA 304, 171, and $131\,\mathrm{\AA}$ (Figs.~\ref{fig:overview}(b)--(d)), and the prominence bubble and plumes are shielded in these channels by the PCTR. The prominence threads are dark in AIA 193 ($\sim 1.6\,\mathrm{MK}$, Fig.~\ref{fig:overview}(e)) and $211\,\mathrm{\AA}$ ($\sim 2.0\,\mathrm{MK}$, Fig.~\ref{fig:overview}(f)) due to continuum absorption, and the bubble and plumes are visible as bright features compared to prominence threads. The three NVST H$\mathrm{\alpha}$ spectral maps in Figs.~\ref{fig:overview}(g)--(i) are shown in the same brightness scale for estimating the Doppler shifts. The bright bubble--prominence interface and dark plumes are clear at H$\mathrm{\alpha}$ $-0.4\,\mathrm{\AA}$ and center.

The formation and evolution processes of the plumes that we focus on are shown in Fig.~\ref{fig:evol}. The formation of the plumes starts from the elevation of the bubble--prominence interface, and the interface gets brighter simultaneously (Figs.~\ref{fig:evol}(a)--(b)). The rising interface is transformed into two plumes with a finger--shaped feature formed (Fig.~\ref{fig:evol}(c)). Then the two plumes continue rising at a speed of $\sim 14\,\mathrm{km\,s^{-1}}$, and the finger gets longer (Fig.~\ref{fig:evol}(d)--(e)). During the evolution of plumes, shorter and denser fingers occur firstly at the bubble--prominence interface (Fig.~\ref{fig:evol}(e)), then move into the plume (Fig.~\ref{fig:evol}(f)). At the late phase of the plume evolution, more fingers occur along the plume boundary (Figs.~\ref{fig:evol}(g)--(h)). The plane--of--sky (POS) velocities calculated using the FLCT technique show obvious flows along the prominence boundary, including the region where fingers occur (Fig.~\ref{fig:evol}(h)). The rising plumes incline leftwards, which is consistent with the flow direction. In addition, a small plume appears at lower height (Fig.~\ref{fig:evol}(f)), which is also split and fingers occur between plumes (Figs.~\ref{fig:evol}(g)--(h)) as the large plumes mentioned before. The finger structures have been predicted in RT instability simulations \citep{2011ApJ...736L...1H,2015ApJ...806L..13K,2016ApJ...825L..29X} and the nearly horizontal fingers were reported by \citet{2019FrP.....7..218A}.

\subsection{Flows along the prominence boundary}
\begin{figure}
   \centering
   \includegraphics[width=5.7in, angle=0]{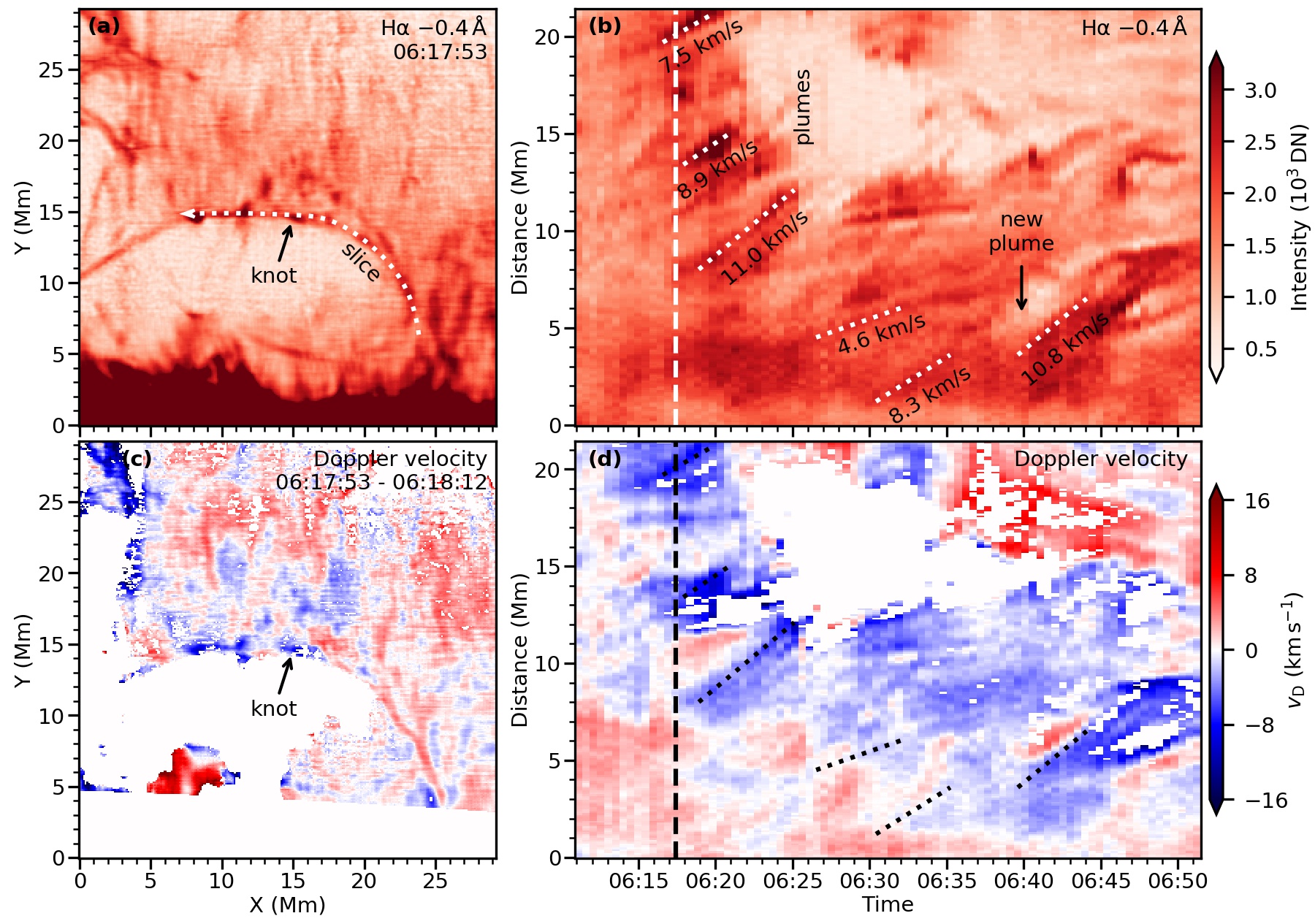}
   \caption{\label{fig:flow}
   Flows along the bubble--prominence interface. Left column: H$\mathrm{\alpha}$ $-0.4\,\mathrm{\AA}$ (a) and Doppler velocity (c) maps before the plume formation. Right column: Time--distance diagrams along the slice marked in (a). Trajectories of some flows are marked in (b) and (d) with dotted lines. The vertical dashed line marks the observation time of (a). Panels (a) and (b) are shown in saturated and reverse scale.}
\end{figure}

We find that the flows along the bubble--prominence boundary (Fig.~\ref{fig:evol}(h)) already exist before the plume formation, which are obvious in H$\mathrm{\alpha}$ $-0.4\,\mathrm{\AA}$ images. In the reverse and saturated H$\mathrm{\alpha}$ $-0.4\,\mathrm{\AA}$ map in Fig.~\ref{fig:flow}(a), some bright knots appear along the prominence boundary, which are blue shifted (Fig.~\ref{fig:flow}(c)). We synthesized time--distance diagrams along the bubble--prominence boundary (the slice in Fig.~\ref{fig:flow}(a) from bottom--right side to the top--left side of the bubble), and the H$\mathrm{\alpha}$ $-0.4\,\mathrm{\AA}$ and Doppler velocity diagrams are shown in Figs.~\ref{fig:flow}(b) and (d), respectively. The top part of the slice misses the prominence at $\sim$06:25~UT, which is due to the elevation of the prominence during the plume formation. Obvious flows occur around 06:17~UT, $\sim10$ minutes before the plume formation. Almost all the knots flow from the lower right side to the upper left side, and are blue shifted. Such motions continue till the end of the diagram, including the period during which the new plume occurs.  Their speeds are non--uniform, generally within $12\,\mathrm{km\,s^{-1}}$ on POS (Fig.~\ref{fig:flow}(b)) and within $8\,\mathrm{km\,s^{-1}}$ along LOS (Fig.~\ref{fig:flow}(d)). A typical knot is marked in Figs.~\ref{fig:flow}(a) and (c). Its POS velocity is $v_\mathrm{POS}=8.9\pm0.8\,\mathrm{km\,s^{-1}}$ and the Doppler velocity is $v_\mathrm{LOS}=-6.8\pm5.0\,\mathrm{km\,s^{-1}}$ (toward us). Therefore, the total speed is $v_\mathrm{total}=11.2\pm 3.1\,\mathrm{km\,s^{-1}}$, and the angle with POS is $37\degr \pm 20\degr$.

\subsection{Turbulence at plume fronts} \label{subs:tur}
\begin{figure}
   \centering
   \includegraphics[width=5.7in, angle=0]{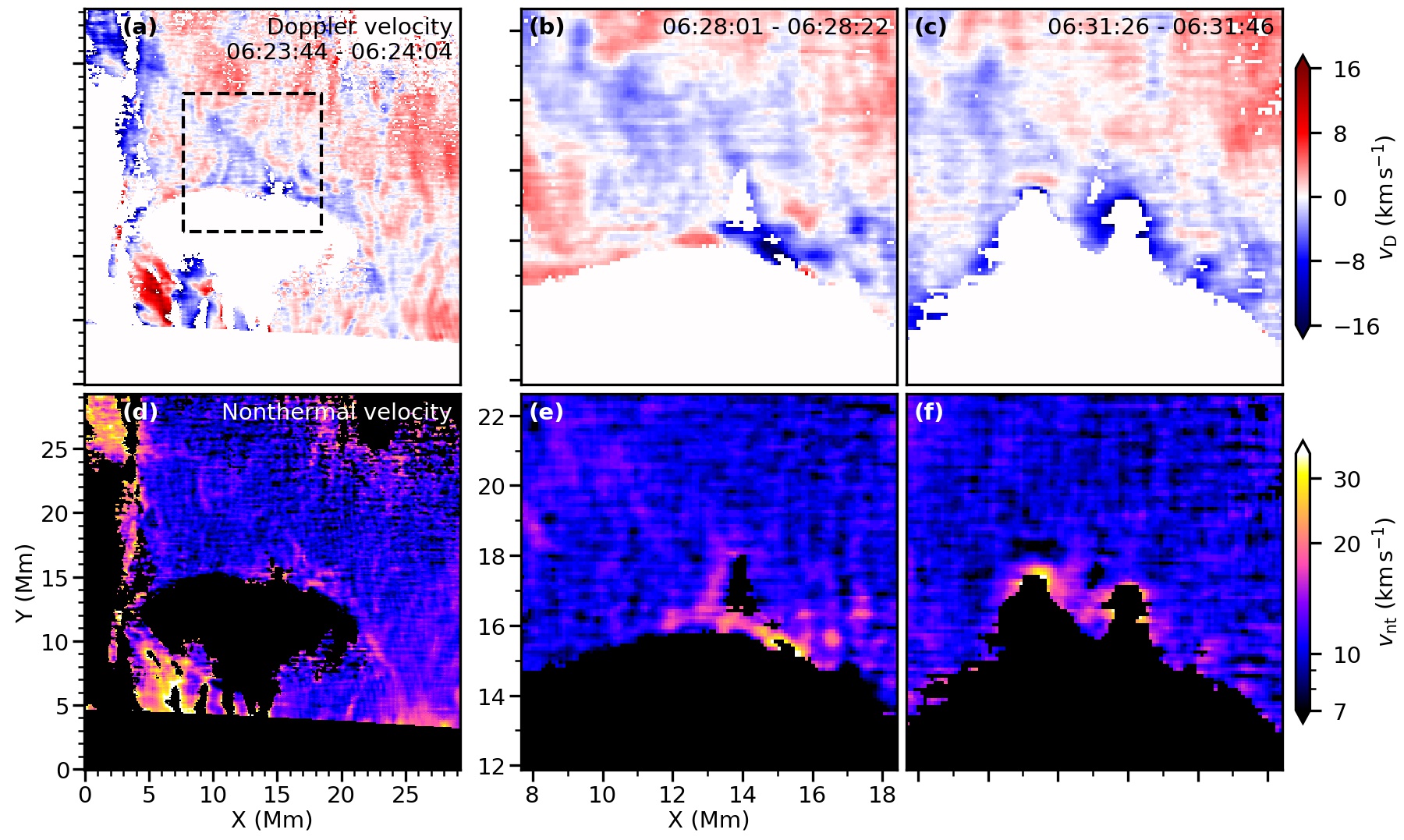}
   \caption{\label{fig:turb}
   Turbulence at plume fronts. (a)--(c) Doppler velocity maps. (d)--(f) Nonthermal velocity maps in logarithmic scale. The dashed black square in (a) shows the FOV of (b)--(c) and (e)--(f).}
\end{figure}

\begin{figure}
   \centering
   \includegraphics[width=5.7in, angle=0]{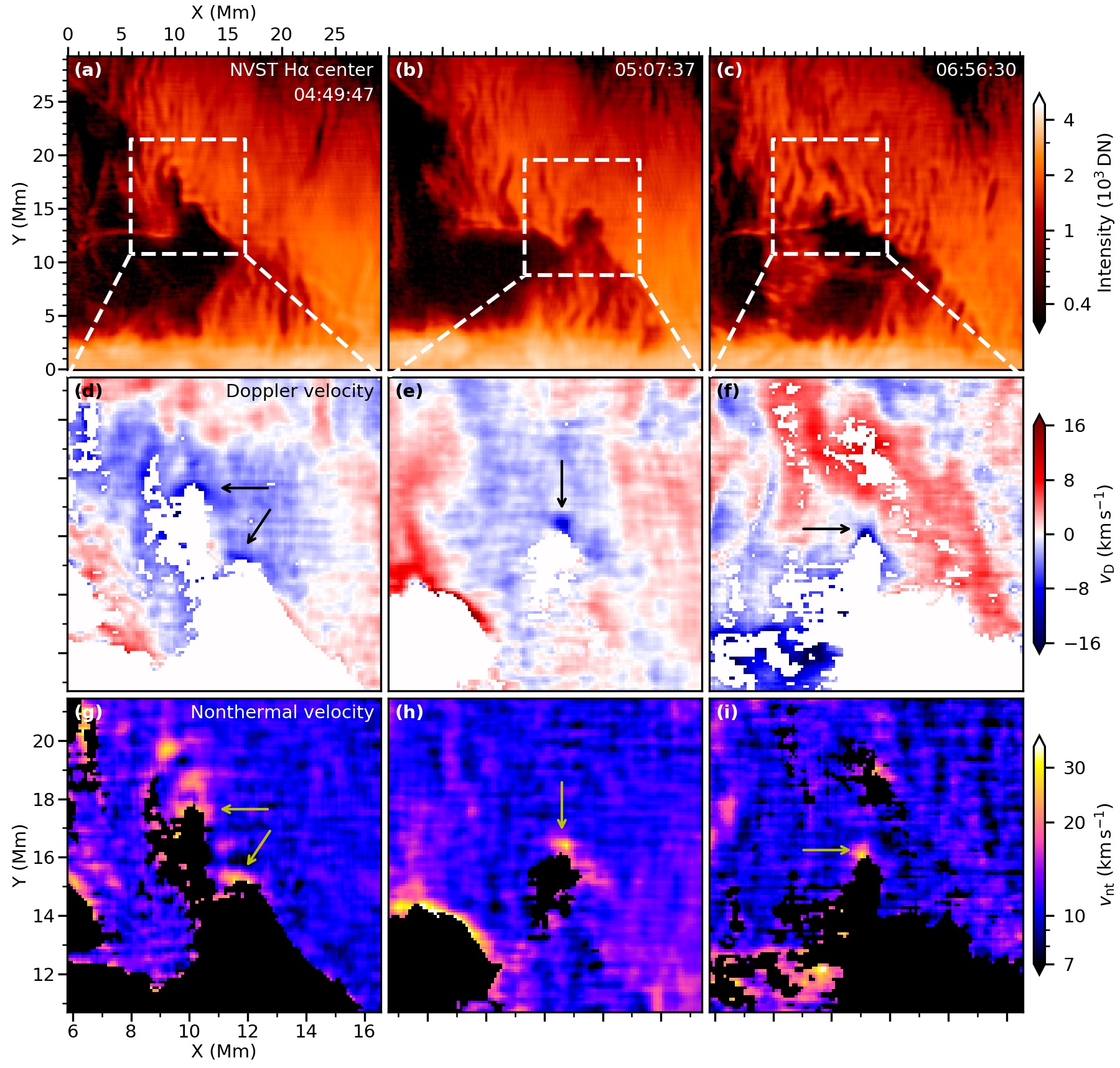}
   \caption{\label{fig:other}
   Other plumes showing brightening, blue shifts, and large nonthermal velocities at their fronts. (a)--(c) H$\mathrm{\alpha}$ center maps. (d)--(f) Doppler velocity maps. (g)--(i) Nonthermal velocity maps. The white squares in (a)--(c) show the FOV of the lower panels. H$\mathrm{\alpha}$ center and nonthermal velocity maps are shown in logarithmic scale.}
\end{figure}
Figure~\ref{fig:evol} shows the brightening enhancement at plume fronts during their formation. Their $v_\mathrm{D}$ and $v_\mathrm{nt}$ maps are plotted in Fig.~\ref{fig:turb}. Compared with the initial maps (Figs.~\ref{fig:turb}(a) and (d)), $v_\mathrm{D}$ at the plume fronts increases significantly and reaches $-16\pm 10\,\mathrm{km\,s^{-1}}$. At the same time, the region also gets more turbulent and $v_\mathrm{nt}$ is $>26\pm 10\,\mathrm{km\,s^{-1}}$ after subtracting the instrumental broadening ($(c/\lambda_0)w_\mathrm{instr}$ is expected to be $<10\,\mathrm{km\,s^{-1}}$ from Figs.~\ref{fig:turb}(d)--(f)). When plasma temperature is in the range of $6000 - 15\,000\,\mathrm{K}$, the corresponding sound speed is $\lesssim 13-20\,\mathrm{km\,s^{-1}}$. Hence, the large $v_\mathrm{nt}$ already exceeds the local sound speed.

We check other three plume cases in Fig.~\ref{fig:other} and find that the enhancements of brightening, blue shifts, and turbulence commonly occur at the plume fronts. The regions of interest are marked with arrows. In the first event (left column), the nearby two plumes are relatively large and the obvious blue shifts and large $v_\mathrm{nt}$ are distributed along the plume fronts. In the latter two cases (middle and right columns), the plumes are relatively small and the turbulent regions are more compact.

\section{Discussion} \label{sect:dis}
Our observations are consistent with previous plume observations \citep{2010ApJ...716.1288B,2017ApJ...850...60B,2019FrP.....7..218A} but reveal more details due to the high--quality H$\mathrm{\alpha}$ images and spectroscopic analysis. (1) The splittings of some large plumes may more or less explain the formation of vertical threads in prominence, which has been predicted by the MHD simulation of RT instability \citep{2016ApJ...825L..29X}. (2) Flows along the prominence boundary occur before the plumes formation \citep{2017ApJ...850...60B}, and continue during the plumes evolution \citep{2019FrP.....7..218A}. (3) Brightening, blue shifts, and large turbulence widely occur at plume fronts. Through observations and simulations, RT and KH instabilities have been proposed to be important for plumes formation (Section~\ref{sect:intro}). However, our observations may suggest that extra mechanisms beyond KH/RT instabilities are necessary. In the following, we will discuss the possible formation mechanisms of plumes according to our observations.

\citet{2017ApJ...850...60B} found shear flows at the bubble--prominence interface, and proved in theory that the shear flows could enhance the growth rate of coupled KH/RT instabilities, hence contribute to the formation of plumes \citep{2010ApJ...716.1288B,2011ApJ...736L...1H,2015ApJ...806L..13K,2016ApJ...825L..29X}. In the late phase of a plume, \citet{2019FrP.....7..218A} found flows along the plume boundary where fingers occur. In our observations, we found flows along the bubble--prominence boundary before the formation of the plumes that we focused on and the following small plume. The flows exist continuously till the late phases of the plumes. The flows are blue shifted, which is determined by the flows direction and prominence magnetic topology, and blue shifts are also observed at fronts of all the plumes that we studied. It may suggest that all the plumes are related to the flows. The flows may not only be responsible for the formation of plumes, but also drive their evolution. The rising plumes incline leftwards, which is consistent with the direction of flows. Besides, the fingers occur where the flows are strong (Fig.~\ref{fig:evol}).

The RT instability has been very successful in explaining the formation of plumes \citep{2018RvMPP...2....1H}. In our observations, the splittings of plumes and occurrence of fingers provide strong evidence of RT instability. Except for the effect of flows on the plume evolution, the plumes in our observation are similar to those simulated using magnetic RT instability \citep{2011ApJ...736L...1H,2012ApJ...761..106H,2015ApJ...806L..13K,2016ApJ...825L..29X}. However, the prominence main body is stable during the plume evolution, which is different from those simulations, where heavy prominence mass drops down. Besides, no sign of brightening (density enhancement) was found at plume fronts in those works. Actually, the RT instability occurs not only when heavy fluid is over the light one in gravitational field, but also when one fluid penetrates into the other one. The RT instability due to the penetration of light plasma into the heavy plasma was simulated by \citet{2014ApJ...796L..29G}. The enhancements of brightening, Doppler shifts, turbulence, and the splittings of plumes are likely to be caused by the collision between the rising plumes and the prominence. In this process, rising plumes compress prominence mass, and prominence plasma gets more turbulent due to pushing and friction of plumes acting on the prominence. Hence we conjecture that the RT instability is mainly driven by the rising plumes and the flows along bubble--prominence boundary but not gravity. This interpretation permits the global stability of the prominence.

However, in addition to the effect that the flows could increase the growth rate of KH/RT instabilities, it is not clear how much the flows dominate the plume formation. We suspect that flows alone cannot trigger plumes. One reason is that the flows are ongoing for a while before plumes form, and the other evidence is that plumes are generally localized and do not fill the bubble--prominence boundary where flows occur. Present observations show that plumes rise in prominence without obvious deceleration at initial phases \citep{2010ApJ...716.1288B}. If plumes are caused by low--density structures penetrating into prominence, upward forces are necessary to balance the gravity and drag force. Thus some extra mechanisms are necessary to trigger and drive plumes. A possible mechanism is the upward magnetic pressure \citep{2008ApJ...676L..89B}, and random disturbance at bubble--prominence boundary is also necessary to explain the occasional occurrence of plumes.

\section{Conclusions} \label{sect:con}
We observed a number of plumes in a quiescent prominence using high--quality H$\mathrm{\alpha}$ images from the NVST and EUV images from the SDO/AIA, which allowed us to study the formation and evolution of plumes in detail, and explore their common properties. Using the 3-points method, we derived Doppler velocity and nonthermal velocity of prominence plasma from H$\mathrm{\alpha}$ spectral images. The plume formation starts from the elevation of the bubble--prominence interface. Meanwhile, the emission, blue shifts, and turbulence at the interface are enhanced. Some plumes split into small ones during their rising with fingers formed between them. Blue--shifted flows along the bubble--prominence interface are found before and during the plume formation. Small and dense fingers appear during the evolution of plumes where the flows are strong.

Common blue shifts at the plume fronts suggest a close connection between the flows and the plume formation. \citet{2017ApJ...850...60B} proposed that shear flows along the bubble--prominence interface could increase the growth rate of coupled KH/RT instabilities. The plume splittings and the formation of fingers are strong evidence of RT instability. Therefore, our observations confirm the relationship between plumes and KH/RT instabilities. However, we suspect that the RT instability is mainly driven by the rising plumes but not the opposite. The enhancements of emission, Doppler shifts, and turbulence at plume fronts are likely due to the push and compression of rising plumes on the prominence plasma. Hence extra mechanisms in addition to KH/RT instabilities are necessary. A candidate is upward magnetic pressure.

\begin{acknowledgements}
We thank the NVST and AIA teams for providing the data. The NVST is a ground-based telescope in the Fuxian Solar Observatory, Yunnan Astronomical Observatory, Chinese Academy of Sciences. AIA is an instrument on board the SDO, a mission for NASA's Living with a Star program. This work is supported by the National Natural Science Foundation of China (NSFC, Grant Nos. 11427803, U1731241, U1631242 and 11820101002) and by CAS Strategic Pioneer Program on Space Science, Grant No. XDA15052200, XDA15320103, XDA15320300 and  XDA15320301. Jean-Claude Vial gratefully acknowledges the support of Purple Mountain Observatory for his visit in November 2019. Yang Su acknowledges the Jiangsu Double Innovation Plan. Zhi Xu acknowledges the NSFC (Grant No. 11873091) and Yunnan Province Basic Research Plan (No. 2019FA001).
\end{acknowledgements}

\section*{Supplementary movie}
Evolution of the prominence between 06:00~UT and 07:30~UT on 2018 November 10. From top left to bottom right: maps of AIA $193\,\mathrm{\AA}$, NVST H$\mathrm{\alpha}$ $-0.4\,\mathrm{\AA}$, Doppler velocity, H$\mathrm{\alpha}$ center, H$\mathrm{\alpha}$ $+0.4\,\mathrm{\AA}$, and Nonthermal velocity. The dotted squares in AIA $193\,\mathrm{\AA}$ maps mark the FOV of NVST maps, which are the same as the maps in Figs.~\ref{fig:overview}(g)--(i). The supplementary movie is available at http://www.raa-journal.org/docs/Supp/ms4896mov.mp4.

\bibliography{plume_prom}{}

\begin{thebibliography}{30}
\providecommand\natexlab[1]{#1}
\providecommand\JournalTitle[1]{#1}

\bibitem[{Awasthi} \& {Liu}(2019)]{2019FrP.....7..218A}
{Awasthi}, A.~K., \& {Liu}, R. 2019, Frontiers in Physics, 7, 218

\bibitem[{Berger} {et~al.}(2008)]{2008ApJ...676L..89B}
{Berger}, T.~E., {Shine}, R.~A., {Slater}, G.~L., {et~al.} 2008, \apjl, 676,
  L89

\bibitem[{Berger} {et~al.}(2010)]{2010ApJ...716.1288B}
{Berger}, T.~E., {Slater}, G., {Hurlburt}, N., {et~al.} 2010, \apj, 716, 1288

\bibitem[{Berger} {et~al.}(2017)]{2017ApJ...850...60B}
{Berger}, T., {Hillier}, A., \& {Liu}, W. 2017, \apj, 850, 60

\bibitem[{Berger} {et~al.}(2011)]{2011Natur.472..197B}
{Berger}, T., {Testa}, P., {Hillier}, A., {et~al.} 2011, \nat, 472, 197

\bibitem[{Brault}(1978)]{1978fsoo.conf...33B}
{Brault}, J.~W. 1978, in Future solar optical observations needs and
  constraints, ed. G.~{Godoli}, Vol. 106, 33

\bibitem[Chae {et~al.}(2013)]{chae_park_song_2013}
Chae, J., Park, H., \& Song, D. 2013, Proceedings of the International
  Astronomical Union, 8, 85

\bibitem[{Dud{\'\i}k} {et~al.}(2012)]{2012ApJ...761....9D}
{Dud{\'\i}k}, J., {Aulanier}, G., {Schmieder}, B., {Zapi{\'o}r}, M., \&
  {Heinzel}, P. 2012, \apj, 761, 9

\bibitem[{Fisher} \& {Welsch}(2008)]{2008ASPC..383..373F}
{Fisher}, G.~H., \& {Welsch}, B.~T. 2008, in Astronomical Society of the
  Pacific Conference Series, Vol. 383, Subsurface and Atmospheric Influences on
  Solar Activity, ed. R.~{Howe}, R.~W. {Komm}, K.~S. {Balasubramaniam}, \&
  G.~J.~D. {Petrie}, 373

\bibitem[{Gibson}(2018)]{2018LRSP...15....7G}
{Gibson}, S.~E. 2018, Living Reviews in Solar Physics, 15, 7

\bibitem[{Gouttebroze} {et~al.}(1993)]{1993A&AS...99..513G}
{Gouttebroze}, P., {Heinzel}, P., \& {Vial}, J.~C. 1993, \aaps, 99, 513

\bibitem[{Gun{\'a}r} {et~al.}(2018)]{2018ApJ...867..115G}
{Gun{\'a}r}, S., {Dud{\'\i}k}, J., {Aulanier}, G., {Schmieder}, B., \&
  {Heinzel}, P. 2018, \apj, 867, 115

\bibitem[{Gun{\'a}r} {et~al.}(2014)]{2014A&A...567A.123G}
{Gun{\'a}r}, S., {Schwartz}, P., {Dud{\'\i}k}, J., {et~al.} 2014, \aap, 567,
  A123

\bibitem[{Guo} {et~al.}(2014)]{2014ApJ...796L..29G}
{Guo}, L.~J., {Huang}, Y.~M., {Bhattacharjee}, A., \& {Innes}, D.~E. 2014,
  \apjl, 796, L29

\bibitem[{Heinzel} {et~al.}(2015)]{2015ApJ...800L..13H}
{Heinzel}, P., {Schmieder}, B., {Mein}, N., \& {Gun{\'a}r}, S. 2015, \apjl,
  800, L13

\bibitem[{Heinzel} {et~al.}(2014)]{2014A&A...562A.103H}
{Heinzel}, P., {Zapi{\'o}r}, M., {Oliver}, R., \& {Ballester}, J.~L. 2014,
  \aap, 562, A103

\bibitem[{Hillier}(2018)]{2018RvMPP...2....1H}
{Hillier}, A. 2018, Reviews of Modern Plasma Physics, 2, 1

\bibitem[{Hillier} {et~al.}(2012)]{2012ApJ...761..106H}
{Hillier}, A., {Hillier}, R., \& {Tripathi}, D. 2012, \apj, 761, 106

\bibitem[{Hillier} {et~al.}(2011)]{2011ApJ...736L...1H}
{Hillier}, A., {Isobe}, H., {Shibata}, K., \& {Berger}, T. 2011, \apjl, 736, L1

\bibitem[{Keppens} {et~al.}(2015)]{2015ApJ...806L..13K}
{Keppens}, R., {Xia}, C., \& {Porth}, O. 2015, \apjl, 806, L13

\bibitem[{Labrosse} {et~al.}(2010)]{2010SSRv..151..243L}
{Labrosse}, N., {Heinzel}, P., {Vial}, J.~C., {et~al.} 2010, \ssr, 151, 243

\bibitem[{Lemen} {et~al.}(2012)]{2012SoPh..275...17L}
{Lemen}, J.~R., {Title}, A.~M., {Akin}, D.~J., {et~al.} 2012, \solphys, 275, 17

\bibitem[{Li} {et~al.}(2018)]{2018ApJ...863..192L}
{Li}, D., {Shen}, Y., {Ning}, Z., {Zhang}, Q., \& {Zhou}, T. 2018, \apj, 863,
  192

\bibitem[{Liu} {et~al.}(2014)]{2014RAA....14..705L}
{Liu}, Z., {Xu}, J., {Gu}, B.-Z., {et~al.} 2014, Research in Astronomy and
  Astrophysics, 14, 705

\bibitem[{Pesnell} {et~al.}(2012)]{2012SoPh..275....3P}
{Pesnell}, W.~D., {Thompson}, B.~J., \& {Chamberlin}, P.~C. 2012, \solphys,
  275, 3

\bibitem[{Shen} {et~al.}(2015)]{2015ApJ...814L..17S}
{Shen}, Y., {Liu}, Y., {Liu}, Y.~D., {et~al.} 2015, \apjl, 814, L17

\bibitem[{Simon} {et~al.}(1982)]{1982A&A...115..367S}
{Simon}, G., {Mein}, P., {Vial}, J.~C., {Shine}, R.~A., \& {Woodgate}, B.~E.
  1982, \aap, 115, 367

\bibitem[{Vial} \& {Engvold}(2015)]{2015ASSL..415.....V}
{Vial}, J.-C., \& {Engvold}, O. 2015, {Solar Prominences}, Vol. 415

\bibitem[{Wang} {et~al.}(2013)]{2013RAA....13.1240W}
{Wang}, R., {Xu}, Z., {Jin}, Z.-Y., {et~al.} 2013, Research in Astronomy and
  Astrophysics, 13, 1240

\bibitem[{Xia} \& {Keppens}(2016)]{2016ApJ...825L..29X}
{Xia}, C., \& {Keppens}, R. 2016, \apjl, 825, L29

\end{thebibliography}
\bibliographystyle{raa}

\label{lastpage}

\end{document}